\documentclass{llncs}

\usepackage{float}
\usepackage{placeins}
\usepackage{graphicx}
\usepackage{color}
\usepackage{caption}
\usepackage{tikz}
\usepackage{subfig}

\usetikzlibrary{arrows}

\usepackage{pbox}
\usepackage{rotating}
\usepackage{pdflscape}
\usepackage{adjustbox}
\usepackage{cite}
\usepackage{chapterbib}
\usepackage{cleveref}

\usepackage{caption} 
\usepackage{amsmath}

\makeatletter
\renewcommand\tagform@[1]{\maketag@@@ {[\ignorespaces #1\unskip \@@italiccorr ]}}
\makeatother

\usepackage{array}
\newcolumntype{P}[1]{>{\centering\arraybackslash}p{#1}}

\begin{document}

\title{In Quest of Significance: Identifying Types of Twitter Sentiment Events that Predict Spikes in Sales}

\author{Olga Kolchyna\inst{1}, Th\'{a}rsis T. P. Souza\inst{1}, Philip C. Treleaven\inst{1}\inst{2} \and Tomaso Aste\inst{1}\inst{2}}

\authorrunning{Kolchyna et al.} 
%
\tocauthor{Kolchyna et al.}
\institute{Department of Computer Science, UCL, Gower Street, London, UK,\and
Systemic Risk Centre, London School of Economics and Political Sciences, London, UK}

\maketitle

\begin{abstract} 
We study the power of Twitter  events to predict consumer sales events by analysing sales for 75 companies from the retail sector and over 150 million tweets mentioning those companies along with their sentiment. We suggest an approach for events identification on Twitter extending existing methodologies of event study. We also propose a robust method for clustering Twitter events into different types based on their shape, which captures the varying dynamics of information propagation through the social network. We provide empirical evidence that through events differentiation based on their shape we can clearly identify types of Twitter events that have a more significant power to predict spikes in sales than the aggregated Twitter signal. 
\end{abstract}

\smallskip


\newpage

\section{Introduction}

During the last years social-media has been extended beyond its original applications changing the way people communicate, share ideas and opinions. Today consumers leave feedback about their customer experiences and express their views about products on social media websites, at the same time, people who are interested in purchasing a product are going online to read reviews before making their decision. Thus, the choice of buying or not buying something is being greatly affected by other people's feedback. Social  opinions or sentiment have become valuable by themselves, accordingly with Wright \cite{Wright-2009} ``for many businesses online opinion has turned into a kind of virtual currency that can make or break a product in the marketplace''. By expressing their opinions online people set up trends, evaluations and sentiments of the marketplace. 

In this context, the social-media becomes a powerful tool for predictions. It has already been shown that studying collective behaviour in human society through social-media allowed researchers to answer many questions: for example, how will the stock market move\cite{TwitterPredicts} or how the unemployment rates will change \cite{Unemployment-2014}. However, the domain of business application has not been explored in it's full potential. Information about how people speak about different products can be relevant for the companies to predict sales, achieve more insight on inventory management, plan marketing campaigns and adjust the offer in real time. To our best knowledge, there are no big-scale studies that incorporated social-media data into forecasting model for sales of a company.  Some research works in this domain predicted opening sales for movies \cite{MoviesPrediction} and book sales \cite{Gruhl:2005}. However, movies and books are phenomena that receive a lot of social attention. The motivation of our study is to investigate whether sales for the general consumer products that are not "trending" can be predicted using social-media sources, particularly Twitter. 

The biggest challenge in building sales forecasting models is availability of sales data. One of the contributions of our study is that the analysis is done for a large dataset, containing daily sales for 75 brands from the  retail sector, including apparel, shoes, bags, body care, furnishings, fashion and games. The data covers the period of 1 year from November 1, 2013 to October 31, 2014 and was supplied by Certona Inc., a company that provides a personalisation platform for the world's most renown brands. To study the predictive power of Twitter we surveyed over 150 million tweets related to the 75 brands and analysed the sentiments they are expressing (positive, negative or neutral) using our own algorithm for sentiment classification. Our dataset is complete, since it includes all brand-related messages posted on Twitter over the specified period of time.

While predicting the direction of sales time-series is important, our ultimate goal is to predict anomalous increases in sales based on Twitter data. Outbursts of activity on Twitter for a particular brand might be an indication of a growing interest toward the brand's products and potential increase in sales. A spike in sales or, in other words, a sales event might mean an opportunity of profits for a business or lost customers in case, when a company was not ready for the outburst of demand. It is essential for the management and operations team to receive alerts about potential sales events so they could take necessary actions for optimising the stock. 

For studying spikes in sales and Twitter data we leverage the techniques from the event study field. Event study was introduced for the first time  by James Dolley in 1933 \cite{Dolley:1933}, and it is generally used to measure the effects of economic events on the value of the companies. Accordingly with Mitchell and Netter \cite{Mitchell:1994} , ``an event study methodology is a statistical  technique  that  estimates  the  stock  price  impact  of  occurrences  such  as  mergers, earnings announcements, and so forth.'' Some applications in other fields are also present. For example, Schwert \cite{Schwert:1981} showed the impact of changes in regulatory system on the value of a firm, Mitchell and Netter evaluated damages in legal liability events analysis \cite{Mitchell:1994}. However, up to date there are no extensive studies that measure the impact of a specific event on sales of a company. To our knowledge, the only study that mentions evaluation of Twitter events in relation to sales is the study by Dijkman et al. \cite{Dijkman:2015}, however their analysis is limited to just one company and 12,780 tweets. 

Since the publication by Fama et. al in 1969 \cite{Fama:1969}, few modifications were introduced to the technique of event study \cite{Brown:1980},  \cite{Kothari:2005}, \cite{Brown:1985} (for more details on the evolution of the methodology refer to \cite{Corrado-2010}). However, the definition of what is an event remained unchanged up to the present day. As outlined by Mac Kinlay \cite{MacKinlay:1997} to perform  an event study it is necessary to identify the event date and finalise the event window within which the analysis is performed. We propose, that with the advances of social-media networks a definition of an event must be changed. While in pre social-media era the day of the news announcement could be considered as an event, nowadays, when critical information first emerges through Twitter, an event is not just a simple announcement moment, but a much broader phenomena which includes the dynamic of opinion propagation through the social network. It has been shown that the necessary step of defining an event window in itself presents a challenge (\cite{Konchitchki:2011},\cite{Bessembinder:2009}, \cite{Cheng:2007}, \cite{Rubin:2013}, \cite{Barber:1997}, \cite{Brown:1985}). In the situation when the process of news propagation may last from few hours to few days and even weeks, the problem of event window definition becomes even more critical. We argue, that information about the event duration has to be incorporated into study of the event after-effects. In this paper, for the first time in event study,  we define an event not by a single date, but considering other features: duration, the peak, growth and relaxation signatures. We also do not specify an event window, but analyse the predictive power of Twitter along a wide range of windows. The statistical approach that we employ allows us to automatically identify the windows during which Twitter events have significant predictive power.

It has been highlighted by the scholars that identifying different types of news events can shed some light on the problem of identifying the most predictive events (\cite{Sprenger:2014}, \cite{Antweiler+Frank:2004}, \cite{Thompson:1988}). The distinction between the different types of events is done based on the topic that is being discussed: initial public offerings, or earnings announcements, or stock splits. We believe, that in the context of social-media it is important to distinguish different types of events not only based on the content of the discussion, but also to take into account how information spreads through the network. For example, a discussion on Twitter will evolve differently through time depending on whether it was initiated by the brand through a marketing campaign or became a result of a word-of-mouth information sharing. A post-event effect on sales might also be different for the two scenarios. Studying these internal dynamics is a challenging task and  requires understanding of the rules of human collective behaviour. Sornette, Crane  and Sano et. al employed techniques from the complex systems theory and were able to get insights into the nature of spikes in blogosphere activities \cite{Yukie:2013}, views of Youtube videos \cite{Crane:2008p6660}, Amazon books sales \cite{Sornette:2004} by studying the growth and relaxation signatures of those spikes. In this study we analyse different types of Twitter events based on their signatures with the aim to discover the class of events that have the highest predictive power. 
\section{Related Works}

The purpose of this study is to investigate whether Twitter events can be used to predict spikes in sales of retail products.A similar objective was set-up by the researchers at HP \cite{ticketSales}, where they aimed to predict the level of ticket sales for movies based on Twitter information. The team achieved very good results managing to predict the revenue of the opening weekend with 97.3\% accuracy – even higher prediction rate than a known prediction tool for the movies, the Hollywood Stock Exchange. Another attempt to predict sales was performed by Gruhl et al. \cite{Gruhl:2005}. In their analysis they used blog data instead of Twitter. 

As showed by Dijkman et al. \cite{Dijkman:2015} the significant relationship between tweets and sales does not hold for the products that do not receive large social attention as movies and books. In the study by Dijkman et al. \cite{Dijkman:2015} to understand the power of twitter to predict sales the researchers performed a Granger causality test and calculated correlations between sentiment and sales figures. They showed that in order to find significant correlations it is necessary to filter tweets based on their polarity, type of user and type of tweet. They also analysed the predictive power of Twitter spikes to predict twitter events, but their definition of the event has limitations. This study used a small size of the Twitter dataset and the analysis was performed only for one company. In this respect, our study presents a large scale analysis across 75 brands analysing all the tweets that were posted regarding selected brands during the period of one year.

Sprenger et al. \cite{Sprenger:2014} and Ranco et al. \cite{Ranco:2015} recently presented studies on the relationship between Twitter and stock prices. Both studies employ the methodology from the event study to find the correlations between Twitter events and the abnormal market returns. Sprenger et al. \cite{Sprenger:2014} arguers that it is important to take in account multiple types of events. In their study they analyse such categories of events, as Corporate Governance, Financial Issues, Operations, Legal Issues, Technical Trading, etc. They show, that different categories of events have a different effect on the market, for example, events regarding Earnings and Mergers and Acquisitions contain information that moves the market, while events of types Product development and Joint Ventures have no effect on the markets. Ranco et al. \cite{Ranco:2015} differentiated between Earnings Announcements(EA) events and non-EA events and showed that while EA events have a stronger impact on the returns, non-EA events also have a significant relationship with the market. Both Sprenger et al. \cite{Sprenger:2014} and Ranco et al. \cite{Ranco:2015} provided empirical confirmation that Twitter sentiment is a stronger signal than Twitter volume. 

Also in our study we use event analysis and differentiate between different types of events taking into consideration their sentiment. However, there are major differences in the way our study is performed. First of all, our goal is to analyse the effect of Twitter on the consumer sales. Secondly, we propose a new way of identifying the event, that takes in account event duration. We argue, that since events on twitter might last from few hours to few days and selection of the time window is critical, the duration of the event has to be taken into account. We perform event analysis not only for Twitter, but for sales as well. Another difference with the above works lies in the way we assign a sentiment label to the events. Sprenger et al. \cite{Sprenger:2014} and Ranco et al. \cite{Ranco:2015} identify events as spikes in volume and then assign a positive, negative or neutral label to them based on the polarity of tweets for the peak day. Our analysis of tweets' sentiment showed, that the proportion of the positive tweets in a day is much higher than the proportion of the negative and neutral tweets. This relationship holds even during the days when there is a rise of negativity on Twitter. Because of this, many events that have significant increase in negativity will be incorrectly classified as positive events. We believe, that in order to capture the true bursts of positivity on Twitter, one needs to consider the ratio of the number of positive tweets to the number of negative tweets in a day. In our study in order to detect positive events we capture the spikes in this ratio (sentiment ratio) time-series. The spikes of sentiment ratio showed to be different from the spikes in Twitter volume. Finally, our study differs from the works by Sprenger et al. and Ranco et al. in a way we differentiate between different types of events. Both groups of researchers categorise the events based on their content. Springer et al. uses manual labelling of different types of events based on the topic discussed. This could become a limitation when a new topic emerges or when methodology is to be expanded to the other domains, such as sales prediction. Event though, undoubtedly, the information about topic is very important, we believe, that in the context of sales predictions the after-event effects on product demand depend not only on the topic discussed, but on the underlying nature of the event (was it caused by marketing campaign? was it a sale or promotion?). As shown by Sornette et al. \cite{Sornette:2006} the underlying nature can be revealed by analysing the shape of the event.  In our work we perform clustering of different events automatically based on the shapes of the events signatures.

The vast majority of the existing literature focuses on the financial application of the social-media data. The most recent study by Brian Dickinson and Wei Hu, 2015 \cite{Dickinson:2015} presents the results of correlations between sentiment and stock prices of companies. The study shows that a significant correlation is present, however the sign of the correlation might differ for different companies. For example, Microsoft and Walmart showed strong positive correlation, while Goldman Sachs and Cisco Systems showed strong negative correlation. Another recent work by Ekaterina Shabunina, 2015 \cite{Shabunina:2015} presents an opposite result, stating that ``the performance polarity in tweets has less information then the general volume of tweets posted, making  possibly  unnecessary  to  undergo  the  time consuming and demanding task of tweets classification''. Tharsis Souza et al. \cite{Tharsis:2015} obtained results similar to Brian Dickinson and Wei Hu \cite{Dickinson:2015}. In their study they present empirical evidence that Twitter sentiment can be used to predict the excess of log-returns for the stocks of the selected retail companies. They also demonstrated that Twitter sentiment has a stronger relationship with stock returns compared to news, however, a weaker relationship than news for volatility. Benthaus et al. \cite{Benthaus:2015} analysed how the dimension of twitter user influences the financial markets and found no significant impact. This conclusion is opposite to the finding of Yang \cite{Yang:2014} who showed that it is possible to identify a community of Twitter users that  ``provide a better proxy between social sentiment and financial market movement''. A contribution to the field by Zheludev et al. \cite{citeulike:13108056} is identifying a subgroup of securities for which movements in Twitter data can provide valuable information.

There is also a growing body of research that uses Twitter to predict different social phenomena, for example, Antenucci et al. \cite{Unemployment-2014 } predicts unemployment rates using PCA and regression analysis on the Twitter data related to the job losses; Culotta \cite{Culotta:2013}, St Louis and Zorlu \cite{Louis:2012}, Paul and Dredze \cite{Paul:2011}, Aramaki et al. \cite{Aramaki:2011}; Lampos et al. \cite{Lampos:2010} predict influenza infection rates; Bermingham and Smeaton \cite{Bermingham:2011}, Tumasjan et al. \cite{Tumasjan:2010}, Kim and Hovy \cite{Kim:2007} predict election results.

\section{Datasets}

Empirical investigation in the present study was performed by using the following datasets: 

\begin{enumerate}

\item Daily sales data supplied by Certona Inc., related with 75 brands companies over the period of one year, from November 1, 2013, to October 31, 2014. The brands cover the following sectors: apparel, shoes, bags, body care, furnishings, fashion, games.

\item More than 150 million social-media messages surveyed from Twitter over the same time period, that mention the names of selected companies; For the Twitter messages two kind of time-series are created:

\begin{itemize}
\item Daily volume data.
\item Daily sentiment score that is calculated based on the number of positive and negative messages in a day.

\begin{equation}
sentScore=\frac{count(NegSentences_{brand})}{count(PosSentences_{brand})}
\end{equation}

\end{itemize}

\end{enumerate}

\section{Methodology}

\begin{figure*}[t!]
  \centering
	\includegraphics[width=5.0in]{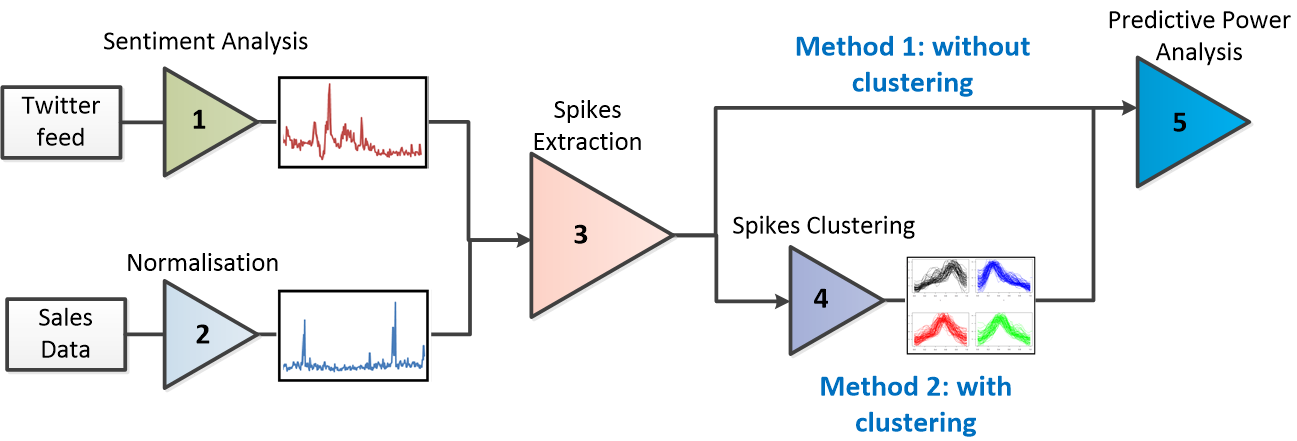}
\caption{\label{fig:methodology} Schematic representation of the methodology for sales spikes prediction.}
\end{figure*}

Schematic representation of our methodology shown in Fig. \ref{fig:methodology}. In the first method we evaluate whether aggregated Twitter events signal can be used to predict sales events. For this purpose we perform the following steps:

\begin{itemize}
\item Twitter sentiment analysis;
\item Events detection;
\item Calculating predictions statistics.
\end{itemize}

In the second method we added an additional step of events clustering in order to find a category of events that have a higher predictive power than the aggregated Twitetr signal. The rational is the following: not all of the events in Twitter have the corresponding event in sales, our aim is to distinguish between the events that have the sale following after them and between all other Twitter events. For this purpose we cluster Twitter events based on their shape and statistically evaluate the predictive power of each Twitter event type. According to the second method the following steps are performed:

\begin{itemize}
\item Twitter sentiment analysis;
\item Events detection;
\item Events clustering;
\item Calculating predictions statistics individually for each Twitter event type.
\end{itemize}

The methodology behind every step for both methods is described in the following subsections.

\subsection{Sentiment Analysis}

Sentiment analysis of Twitter messages was performed using an automated tool that we developed \cite{Kolchyna:2015}. The underlined algorithm was build by combining the traditional lexicon-based approach with the machine-learning algorithm. During the first step of analysis, a lexicon score was calculated using a lexicon of words with assigned to them polarity values. The lexicon was designed specifically for Twitter analysis and apart from traditional sentiment words incorporated Twitter specific emoticons, abbreviations and special grammatical structures that are often used in social-media. During the second step, the support vector machines classifier was trained and the lexicon score calulated during the first step was used as the feature in the classification along with other features, such as number of adjectives, number of negations, number of positive/negative emoticons, etc. The outcome of the algorithm is a label assigned to each Twitter message: positive, negative or neutral. The model was trained and validated on the Twitter dataset from the SemEval-2013 Competition \cite{SemEval-2013}. The results of the validation showed, that our algorithm outperformed all 40 teams that took part in the SemEval-2013 competition by achieving the highest F-score of 73\%. 

\subsection{Normalisation}

Sales data obtained from Certona Inc. was presented in a form of individual transactions with the time stamp and a sales value. We processed the data to create daily figures by summarising the revenues for each day. 

A second step of preprocessing included calculation of the z-score of each point in the data series. This step was necessary for creating comparable values: 

\begin{equation}
\frac{x-\mu}{\sigma}
\end{equation}

\subsection{Events Detection}

\subsubsection{Spikes Identification.}

A Twitter/sales spike is defined as an anomalous deviation from the baseline; an extreme event, outburst of activity on Twitter or outburst in sales. Since the focus of our analysis is to measure whether spikes in Twitter can be used to predict events in sales, the process of spikes detection and extraction is a crucial step. In our analysis we compared the performance of three outlier detection methods:

\begin{enumerate}
\item  \textbf{ESD identifier} (abbreviation for ``Extreme Studentized Deviation'' \cite{Rosner:1983}). This is the most popular outlier detection rule. For a sample $X_N = \left\{x_i\right\}^N_{i=1}$ it classifiers any point more than \textit{t }standard deviations away from the mean to be an outlier, where the threshold value \textit{t} is most commonly taken to be 3.  In other words, $x$ is identified as an outlier if:

\begin{equation}
|x - \overline{X}| \geq t\sigma
\end{equation}

where $\overline{X}$ is the mean and $\sigma$ is estimated standard deviation of the data sequence:

\begin{equation}
\overline{X} = \frac{ \sum^N_{i=1} x_i}{N},    \sigma = \sqrt{\frac{ \sum^N_{i=1}(x_{i} - \bar{X})^2}{n - 1}}
\end{equation}

Motivation for the threshold choice t = 3 comes from the fact that for normally-distributed data, the probability of observing a value more than three standard deviations from the mean is only about 0.3\%. The problem with this outlier detection procedure is that both the mean and the standard deviation are themselves extremely sensitive to the presence of outliers in the data. In fact, if the level of outliers is higher than 10\%, this rule fails completely, detecting no outliers at all. As a consequence, this procedure is likely to miss spikes that are present in the data.

\item \textbf{Hampel identifier} \cite{Hampel:1971}, \cite{Hampel:1974:ICR}. For this outlier detection method, the mean is replaced with median of the residuals and the standard deviation with the MAD scale estimate . MAD is a robust measure of the variability of univariate data. To compute MAD, one calculates the median of the absolute deviations of each historical value from the data's median. 

\begin{equation}
MAD(X_N) = median(|X_1 - median(X_N)|, ..., |X_N - median(X_N)|)
\end{equation}

$x$ is identified as an outlier if:

\begin{equation}
|x - median(X_N)| \geq g(N, \alpha_N)MAD(X_N)
\end{equation}

Where $g$ is a function related to the number of data points and a specified type I error (see \cite{Liu:LiuSJ04}, \cite{Davies:1993} for more details).

The Hampel filter is much more resistant to the influence of outliers, however, it can be too aggressive in classifying values that are not really extremely different and result in many false-positive spikes.  It can be shown that if more than 5\% of the data points have the same value, MAD is computed to be 0, so any value different from the residual median is classified as an outlier. 

\item  \textbf{Median and Interquartile Deviation Range Method (IQR)} \cite{Tukey:1977} CHECK. For this outlier detection method, one calculates the 25th percentile and the 75th percentile of the data. The difference between the 25th and 75th percentile is the interquartile deviation (IQR). The historical value $x$ is classified as an outlier if it is outside of the closed range:

\begin{equation}
[Q_1 - K(IQR);  Q_3 + K(IQR)] 
\end{equation}

where $IQR = Q_3 - Q_1$, where $Q_1$ and $Q_3$ are the 25th and the 75th percentiles respectively, and K is often selected equal to 1.5.

\end{enumerate}

In this paper we first test the three methods and then, adopt the third method for the analytics.

\subsubsection{Accounting for Non-Stationarity and Weekly Patterns.}
Sales and Twitter time-series are non-stationary. To account for non-stationarity the measures of data variance such as “MAD scale estimate”/Standard Deviation/IQR are computed from the observations within a moving window.  The length of the window is selected such, that the data within that window could be considered stationary. 

\begin{figure}[htp] 
\centering
\includegraphics[width=1.0\textwidth]{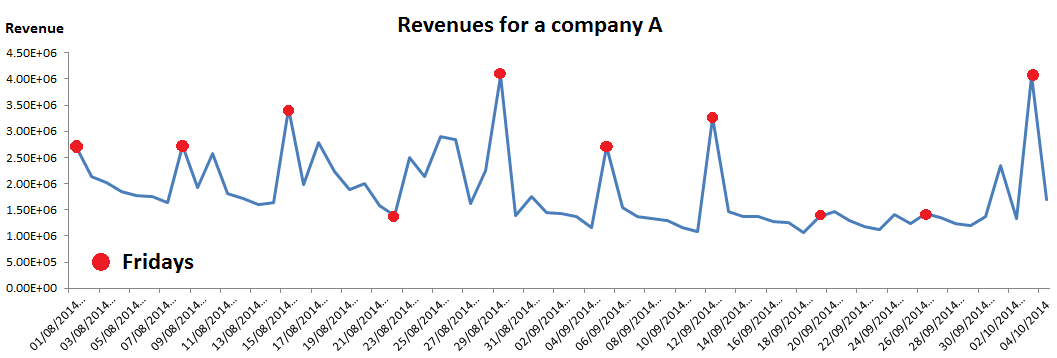}
\caption{\label{fig:frRevenues}Daily revenues for a company A. The red circles indicate weekly Friday sales.}
\end{figure}

We also observe  that sales and Twitter data exhibit weekly patterns. For example, during the week we can see that tweets volume on Fridays and weekends is much higher than during the other days of the week. If Friday tweets volume is to be compared to Thursday tweets volume, the outcome of the spikes identifier would show a spike on Friday, however we want to detect only special events and not regular bursts. For this purpose we would need to compare Friday tweets volume data to the data from the previous Friday. Fig. \ref{fig:frRevenues} presents revenue data for one of the brands under analysis. Red circles in a graph indicate Fridays. As can be seen from Fig. \ref{fig:frRevenues} during most of the Fridays there is a peak in revenues. From the brand information we know that those peaks are related to weekly promotions that the company runs and the spikes in revenue on Fridays should not be considered special events. To account for weekly patterns in the process of spikes detection we construct a moving window that is comprised of data points from the same day of the week. For example, if the data point of primary interest is Friday, a moving window is constructed that includes the data point of primary interest and the K prior Friday values. In this way Friday will only be compared to K previous Fridays, Saturdays will be compared to K previous Saturdays and so on.

\subsubsection{Identifying Events.}
The spike detection methods described previously allow to identify individual spikes in data. However, one event in social-media or sales might last for few hours, days or even weeks and can be comprised on multiple smaller spikes. The aim of our study is to identify bursty events rather than individual spikes. For this purpose, when an event is detected we extract the peak of the event, the growth and the relaxation. These features are required for further analysis.

For the purpose of events detection and its features extraction we propose an algorithm that is based of the following set of rules:

\begin{enumerate}

\item If two spikes are adjacent they are considered to be part of one event.
\item If two or more individual spikes are combined into one event, the spike that reaches the highest value is considered to be the peak of the event.
\item To identify the growth and the relaxation of the event the analysis of the neighbouring left and right data points is performed according to the rules described below.
\item When extracting event growth and relaxation a maximum distance $maxDist$ from the peak is set as a fixed parameter, for example, 14 days. The growth/relaxation cannot continue longer than the established distance  $maxDist$.
\item To extract the growth we analyse the left neighbour of the event peak. If the left neighbour is smaller than the peak it is considered to be part of the growth signature. We then move to the next left neighbour and compare its value to the previous neighbour. If its value is smaller, the point becomes part of the growth signature as well. The process continues while the data exhibits the down trend.
\item The first left neighbour value that is higher than its previous neighbour value is saved as a $minLeftPoint$.
\item If the $minLeftPoint$ point is higher than the median of data and within the specified distance  $maxDist$ from the peak of the event, we continue checking the following left neighbours for the specified in advance number of steps $X$. If within a specified distance  $maxDist$ the value falls below the previous minimum value $minLeftPoint$, the left neighbours are included as part of the growth and the minimum value $minLeftPoint$ is updated. 
\item In case when none of $X$ following left neighbours have value smaller than the saved minimum value, the check stops and the $minLeftPoint$ becomes the first point of the growth. All the points between the first growth point and the peak of the event are called the growth signature of the event.
\item Identical analysis is performed for the right neighbours of the peak to extract the relaxation of the event.
\item The growth/relaxation signatures of different events should not overlap.
\item When the growth and  relaxation are extracted, a peak is recalculated, since extraction of growth/relaxation could introduce a new peak.

\end{enumerate}

\subsection{Events Clustering}

We look for classes of events with similar shapes. Solving events clustering problem is solving a similarity matching problem for the collection of time series representing events. In this study we used KMeans clustering and compared three different ways of calculating the distance between data points. 

\begin{enumerate}

\item \textbf{Euclidean Distance (ED)} is the most used distance function that calculates the similarity between two sequences of the same length by summing the ordered point-to-point distance between them. 

\begin{equation}
d(T, S) = \sqrt{\sum^N_{i=1}{(T_i - S_i)^2}}
\end{equation}

Where $T$ and $S$ are time series of length $n$. This metric is a distance measure, since it obeys to the three fundamental metric properties: non-negativity, symmetry and triangle inequality \cite{Cai:2004}.

\item \textbf{Dynamic Time Warping Distance (DTW)}. While Euclidean distance is a linear map between points, DWT \cite{Keogh:2005} allows non-linear mapping. It is a more robust measure allowing to compare time series of different lengths. Given two time series $T=\{t_1, t_2, ..., t_n\}$ and $S=\{s_1, s_2, ..., s_m\}$ of length $n$ and $m$ respectively, a distance matrix $n*m$ is constructed where each element represents a pairwise distance between points in the two sequences: 

\begin{equation}
distMatrix = \begin{bmatrix}
d(T_1, S_1)& ...& d(T_1, S_m)\\
...& ...& ...\\
d(T_n, S_n)& ...& d(T_n, S_m)
\end{bmatrix}
\end{equation}

The objective of DTW is to find the warping path $W=\{w_1, w_2, ..., w_K \}$ of continuous elements on distMatrix that minimizes the following function:

\begin{equation}
DTW(T,S) = \min\left (\sqrt{\sum^K_{k=1}{w_k}}\right )
\end{equation}
 
\item \textbf{Euclidean Slopes Distance (ESD}. We propose a new way of calculating the distance measure based on the slopes (first derivatives) of the time series. Our algorithm works as follows: 

\begin{itemize}
\item We divide each time series $X_i = {\{(t_i^1, y_i^1), ..., (t_i^n, y_i^n)\}}$ into $K$ number of sequential strips of equal length along the time-axis \cite{Slopes:2005};
\item Slopes $S_i^k$ for each stripe are calculated as 

\begin{equation}
S^k_i = \frac{(y_i^{k+1} - y_i^k)}{\Delta{t}}
\end{equation}

where $(t_i^k, y_i^k)$ and $(t_i^{k+1}, y_i^{k+1})$ are the start end end coordinates for the $k_{th}$ strip of the $i_{th}$ time sequence; 
\item The Euclidean distance computed using as an input the slopes time series.
\item Parameter $K$ is chosen to be equal to one third of the average length of the time series.

\end{itemize}

\end{enumerate}

\subsection{Predictive Power Analysis}

To analyse the power of Twitter events to predict sales events we first plot sales events superimposed over the Twitter events on the time-line. We consider that a Twitter event has a power to predict sales if there is at least one sales event appearing after the beginning of the Twitter event. In the situations when multiple Twitter events are followed by one sales event we make an assumption that all these Twitter events contributed to the appearance of the sale event, and when calculating the statistics we will count them as one Twitter event predicting sales event. This is the most conservative approach, which prevents us from over-counting the number of predictive events, however may result in under-counting them.

\begin{figure*}[t!]
  \centering
	\includegraphics[width=4.9in]{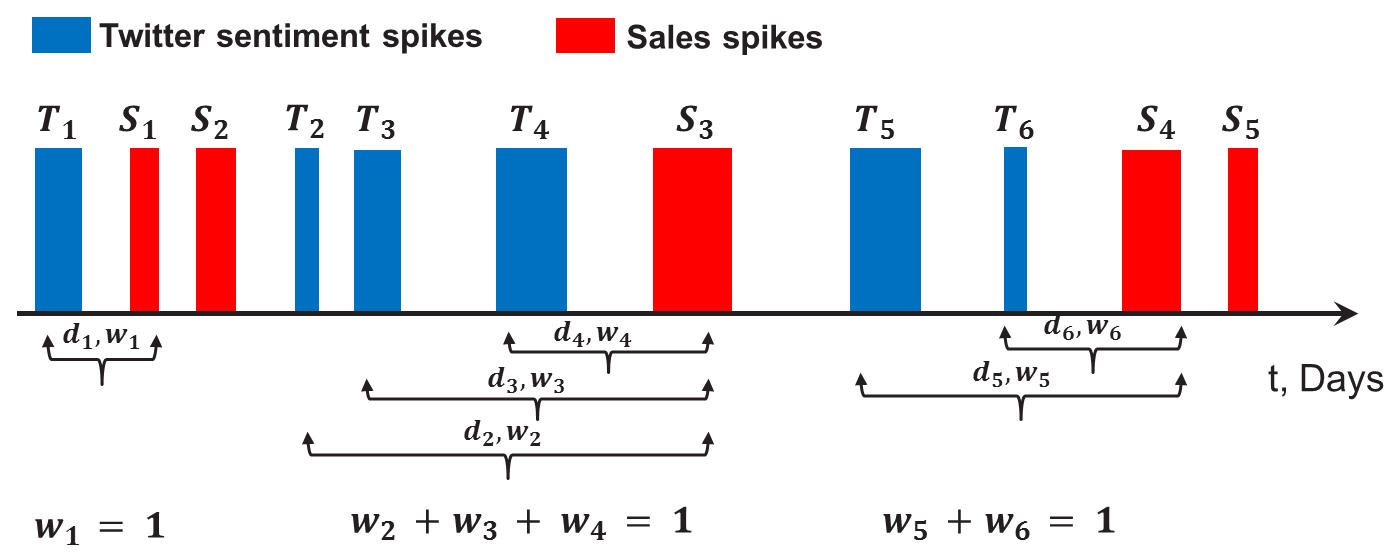}
\caption{\label{fig:spikesWeights}Schematic representation of sales events together with Twitter events. Red bars represent sales events, blue bars represent Twitter events, the width of the bars represents the duration of the event, the difference in the height of the bars represents different events types.}
\end{figure*}

A sales event might happen after a Twitter event at different distances. For each Twitter event we store the distance at which the first sales event happened. We also assign a weight to each Twitter event which is inversely proportional to the distance between the Twitter and the sales events: the longer the distance between sales and Twitter events the smaller is the weight, and vice a versa: the shorter is the distance the higher is the weight. The weights of Twitter events that have one sale following them should sum up to one. For example, in Fig. \ref{fig:spikesWeights} we can see that Twitter events T2, T3 and T4 all have one sale S3 following them. In this case we assume that T2, T3 and T4 all contributed to the appearance of the sales event S3. Each of the Twitter events is assigned a weight w2, w3 and w4 with the sum of the weights equal to one: w2 + w3 + w4 = 1. As mentioned earlier, summing of the weights to one allows to count three Twitter events T2, T3, T4 as one Twitter event and this is a conservative hypothesis. The weights w2, w3, w4 are inversely proportional to the distances d2, d3 and d4. Thus the Twitter event T2 is the furthest from the sale event S3 and will have the smallest weight, while the event T4, being the closest to the sale event, is likely to have the highest contribution to the appearance of the sale event and will have the highest weight.

As the next step we calculate the cumulative probability to have at least one sales event after a Twitter event. This was calculated as follows:

\begin{itemize}
\item Calculate and save the distance in days between the Twitter event T and the first sales event S.
\item Calculate the corresponding weight for each Twitter event.
\item Sort the Twitter events in incremental order of distances.
\item Calculate the probability for each Twitter event to have at least one sale event after it by dividing the total number of events for each distance by the sum of their weights.
\item Compute the cumulative probability for every distance by summing up the probabilities of the previous distances.
\end{itemize}

The cumulative probability is calculated for each distance d and plotted as a graph. This graph allows to see the probability of having at least one sales event after a Twitter event after 1,2,3, etc. days.

\subsection{Statistical validation}

Our null hypothesis, H0, is that Twitter events have no impact on the behavior of sales (sales follow Twitter events in a random manner). We tested our hypothesis using the randomisation test. We reject the null hypothesis if a P-value is less than 0.05. Statistical significance of our test means that the observed number of sales or more are unlikely to occur by chance alone.

We first consider the observed positions of Twitter and sales events (system A) during a year period and calculate the cumulative probabilities of detecting at least one sale event after a Twitter event (as described in the previous subsection). We then randomise the positions of sales events (system B), however, the number and duration of the events are preserved and the randomisation is made in a way that events do not overlap.  The randomisation process is repeated 1000 times. For each run the cumulative probability is calculated. 

To analyse the significance we compute the difference between the observed and randomised cumulative probabilities for each of 1000 runs. We then calculate the average difference, standard errors and confidence interval. A difference below 0.05 is unlikely and thus we can reject a null hypothesis and conclude that the system A has statistically significant advantage over the system B, and the sales are therefore likely to follow Twitter events in a non-random manner. 

\begin{figure*}[t!]
  \centering
	\includegraphics[width=4.9in]{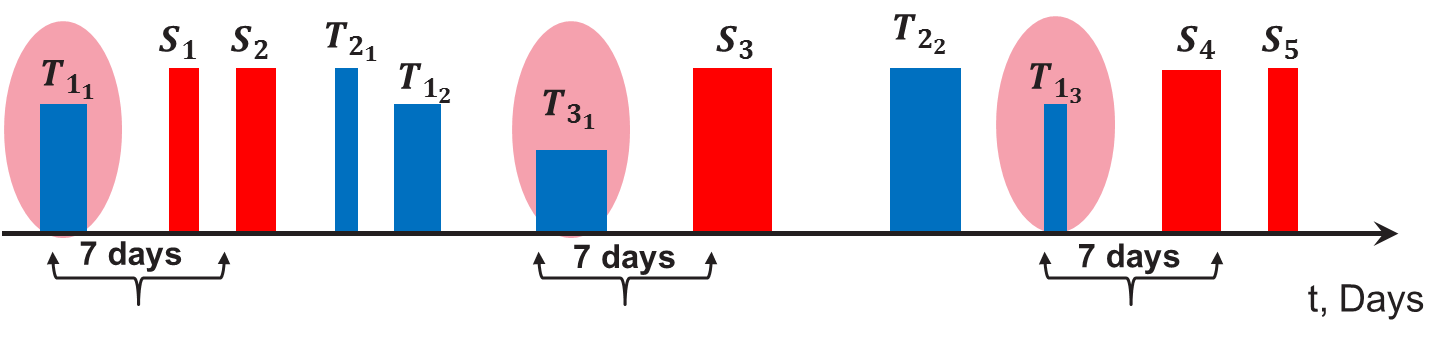}
\caption{\label{fig:sucSpikes}Successful Twitter events highlighted with red circles. A Twitter event is considered to be successful if within a specified time interval (7, 14, 21, etc. days) there is a sales event following.}
\end{figure*}

As a second test we calculate the number of successful Twitter events for the observed and randomised results. A Twitter event is considered to be successful if within a specified time interval (7, 14, 21, etc. days) there is a sales event following (Fig. \ref{fig:sucSpikes}). If the number of the successful Twitter events for the observed results is outside of 95\% confidence interval, we can conclude that the number of Twitter events that predicts events in sales is significant.

\section{Results and Discussion}

\subsection{Events Detection Results}

\begin{figure*}[t!]
  \centering
	\subfloat[]{\includegraphics[height = 1.7in, width=4.5in]{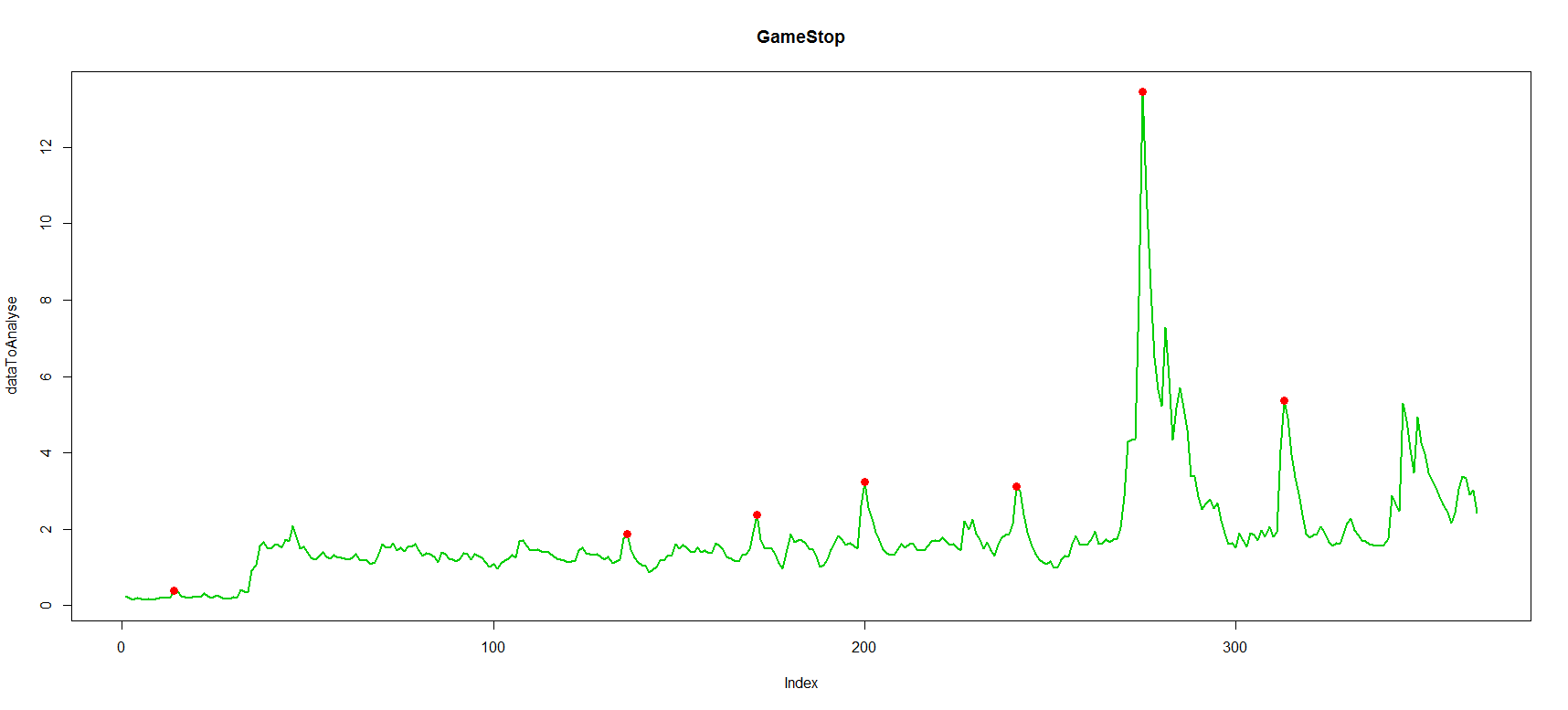}
\label{fig_SDfirst_case}}
\hfil
\subfloat[]{\includegraphics[height = 1.7in, width=4.5in]{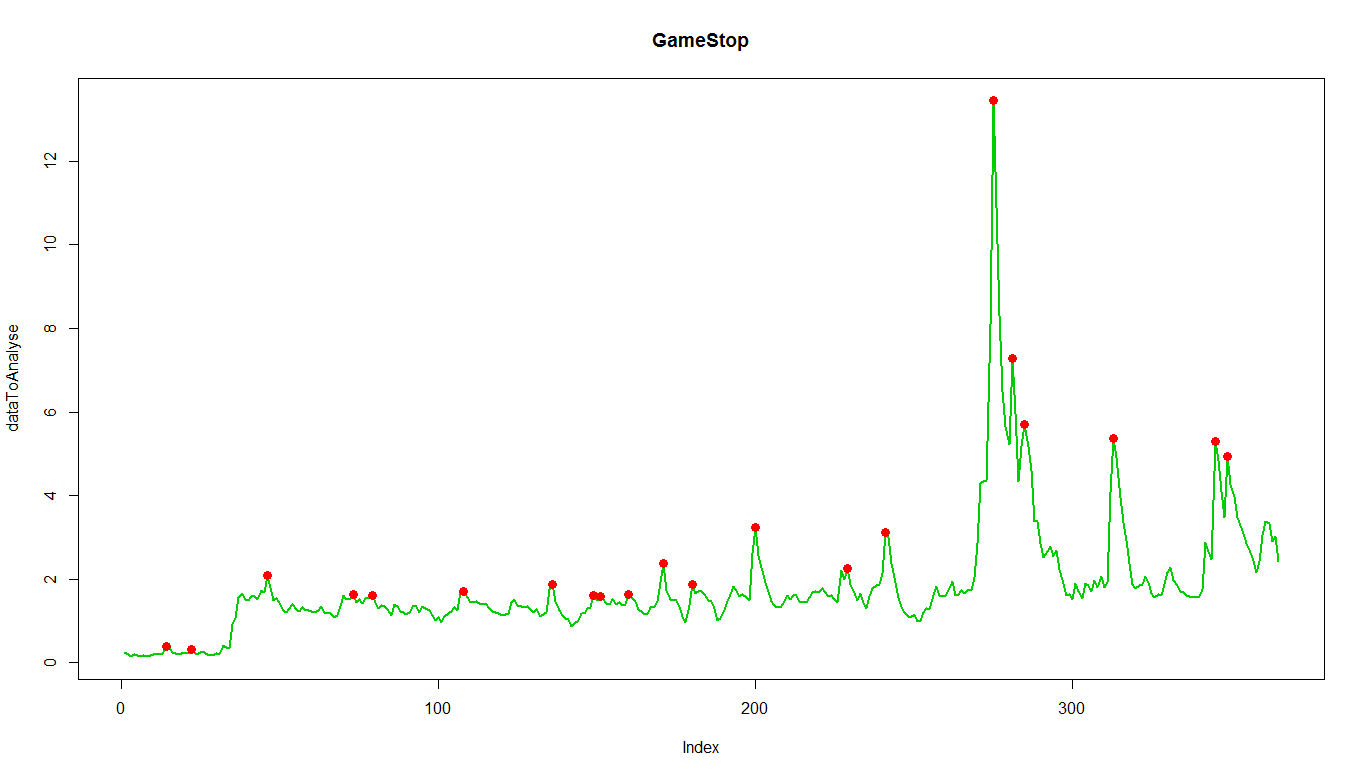}
\label{fig_SDsecond_case}}
\hfil
\subfloat[]{\includegraphics[height = 1.7in, width=4.5in]{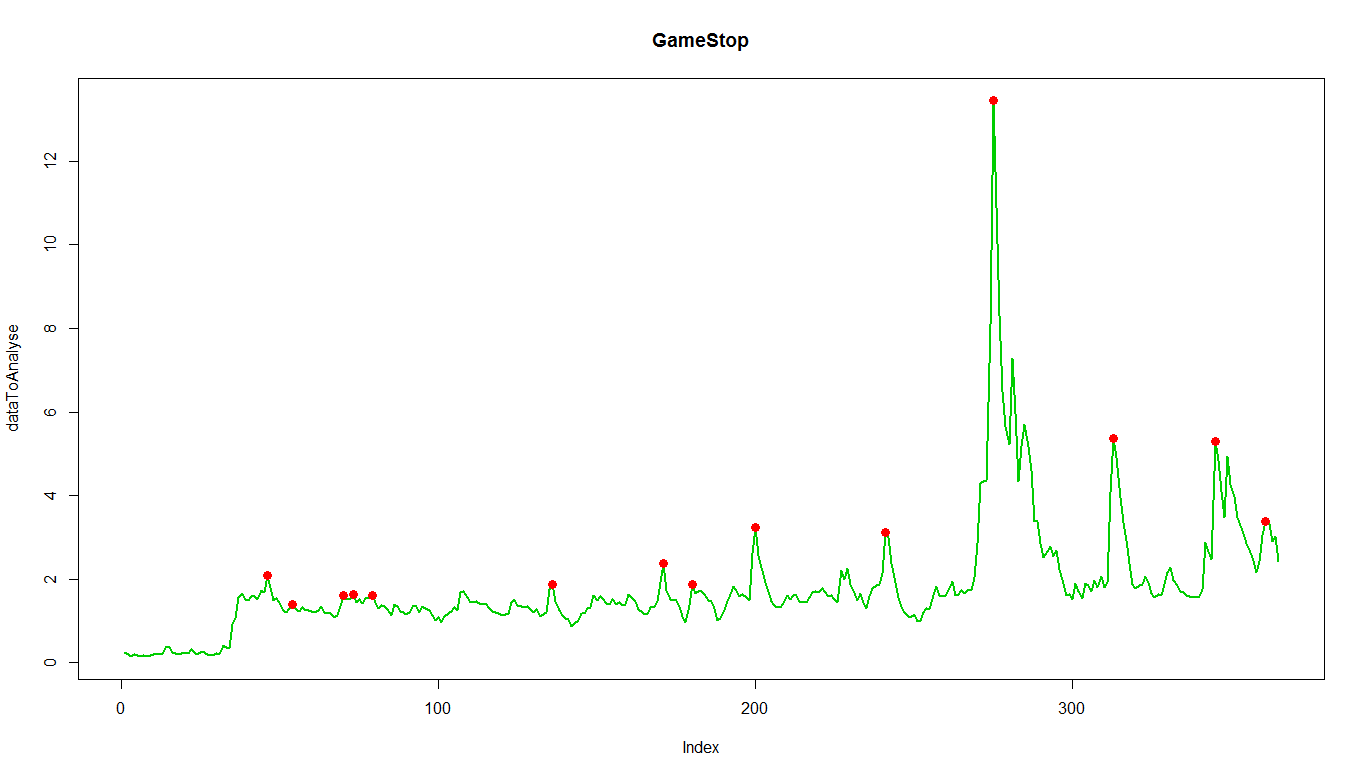}
\label{fig_SDthird_case}}
\caption{Spikes detection for Twitter sentiment time series for a retail brand using (a) ESD identifier, (b) Hampel filter (c) IQR method.}
\label{fig_spikesExtrResults}
\end{figure*}

As described in the Methodology section the first step of events detection process is identifying spikes in time series. In our analysis we used the moving window of 7 days length. The results of spikes extraction process by three algorithms (ESD identifier, Hampel filter, IQR method) for the sales data of one the brands under analysis are depicted in Fig. \ref{fig_spikesExtrResults}.

While ESD identifier missed some of the spikes (a), Hampel filter (b) identified even small increases in sales as spikes. IQR method (c) shows a superior performance comparing to  both ESD and Hampel filter. It captures the big spikes in data and does not suffer from identifying as spikes the non-significant bursts. We therefore use IQR as the method in our final analysis since it provides a good balance between the amount of false-positives and false-negatives.

\begin{figure}[t!] 
\centering
\includegraphics[width=1.0\textwidth]{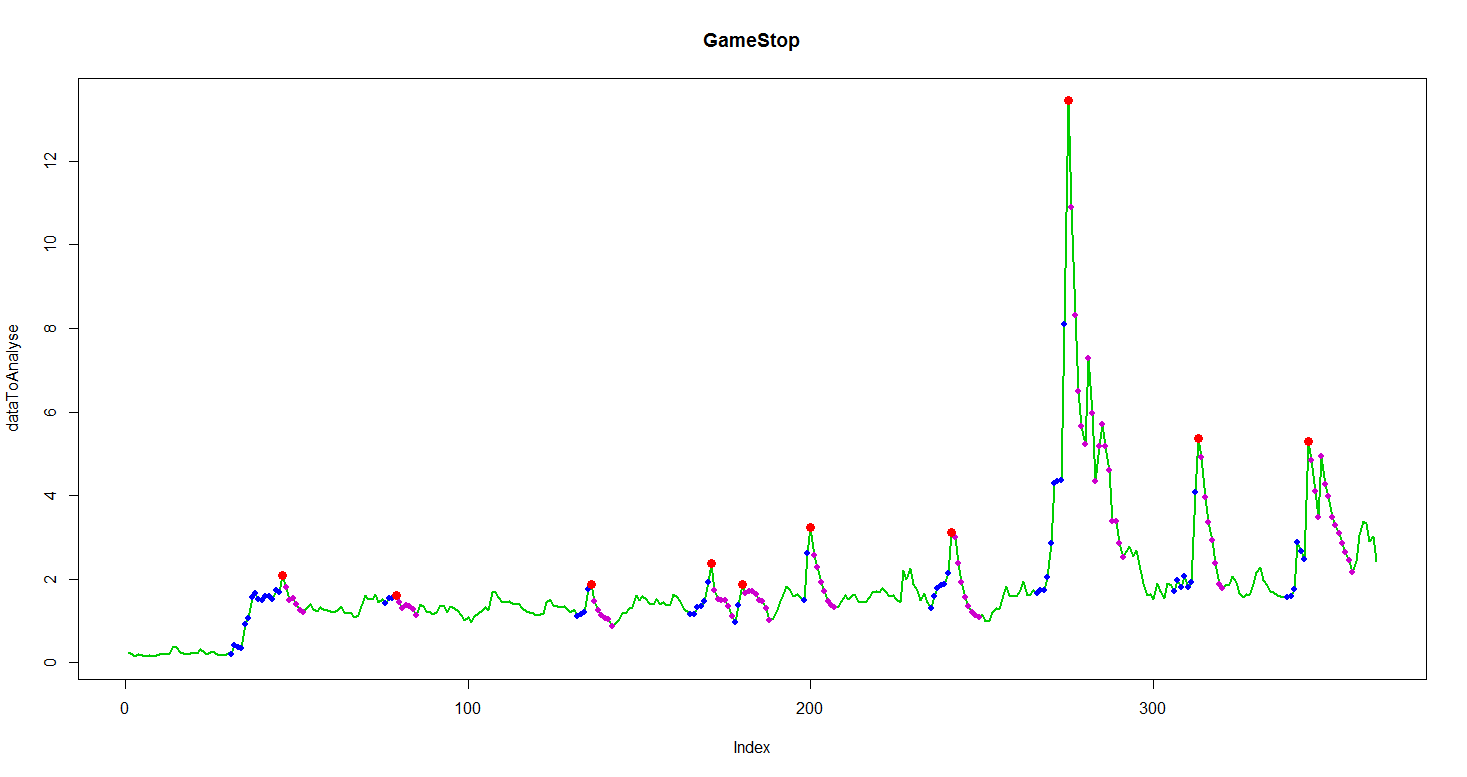}
\caption{\label{fig:spikeSignature}Signatures of Twitter sentiment events for a retail brand. Blue colored dots represent the growth signature of the event, the red circle is the peak of the event and magenta colored dots represent the relaxation signature.}
\end{figure}

In the next step we performed events detection and extraction of events features (peak, growth and relaxation signatures) as described in the Methodology section. As the result we identified 810 spikes in Twitter sentiment data and 760 sales spikes across 65 brands for the period of one year 1 November 2013 - 31 October 2014. Fig. \ref{fig:spikeSignature} demonstrates the outcomes of the event detection algorithm: the events of interest are highlighted with color: the green is the growth signature of the event, the red circle is the peak of the event and the blue is the relaxation signature. 

\subsection{Clustering results}

\begin{figure*}[t!]
  \centering
	\subfloat[Euclidean Distance]{\includegraphics[width=2.3in]{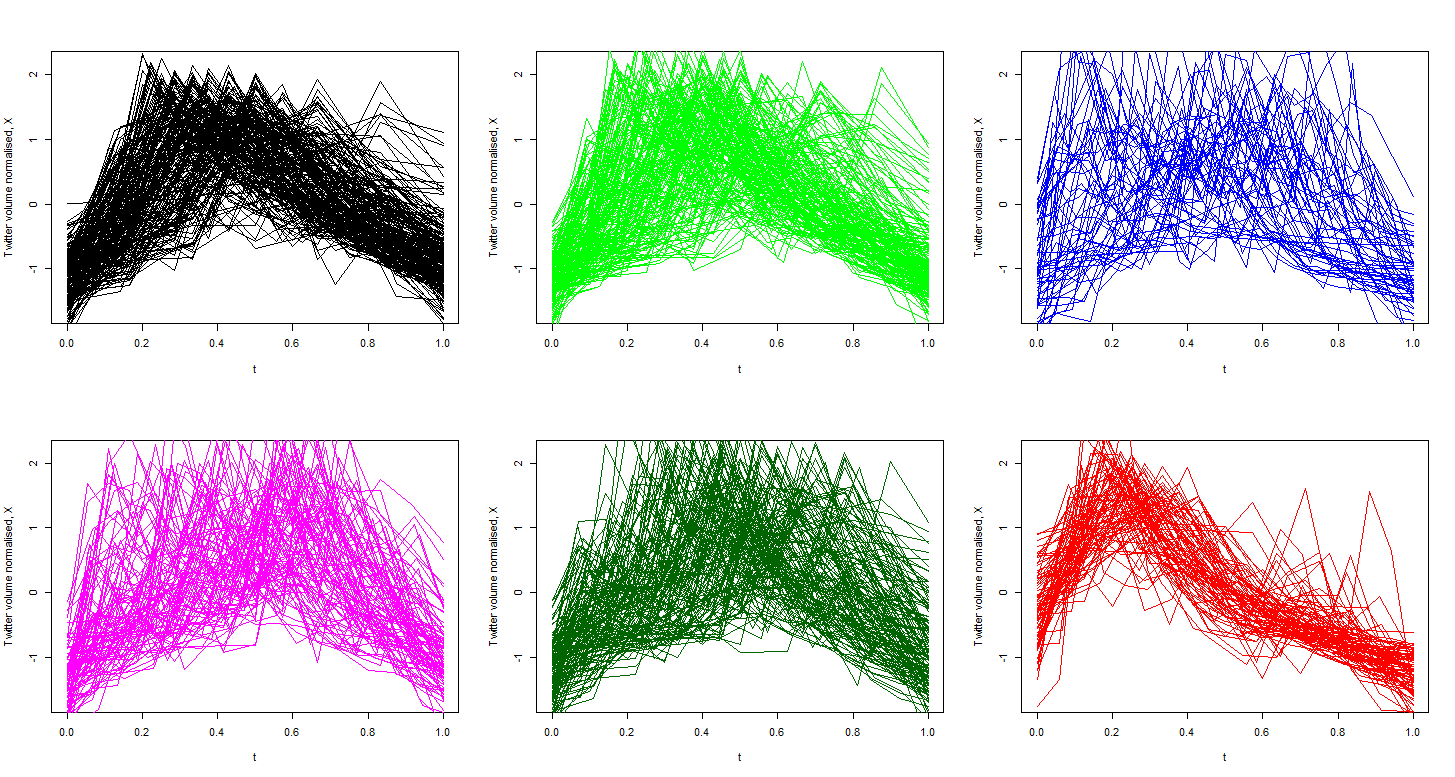}
\label{fig_first_case}}
\hfil
\subfloat[Dynamic Time Warping]{\includegraphics[width=2.3in]{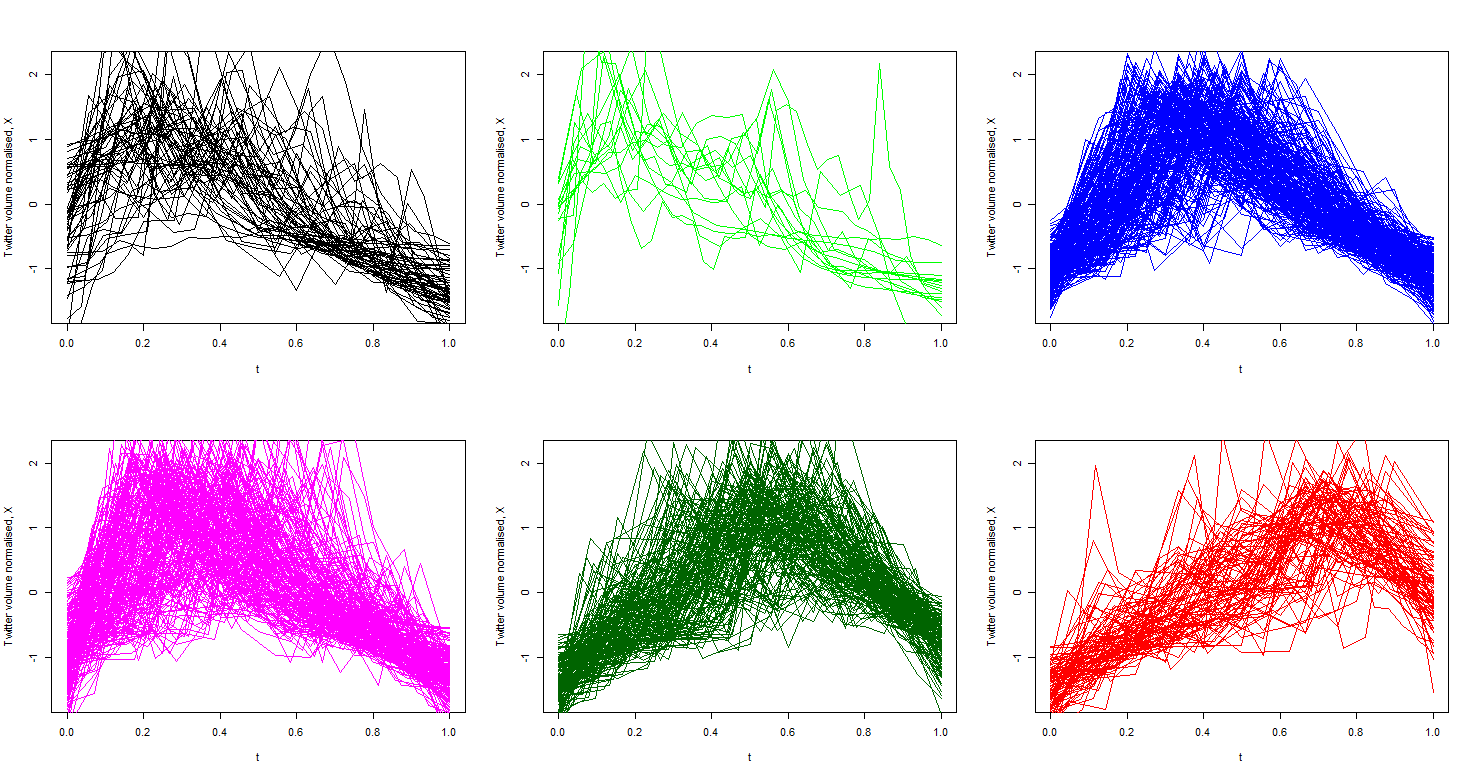}
\label{fig_second_case}}
\hfil
\subfloat[Euclidean Slopes Distance]{\includegraphics[width=2.3in]{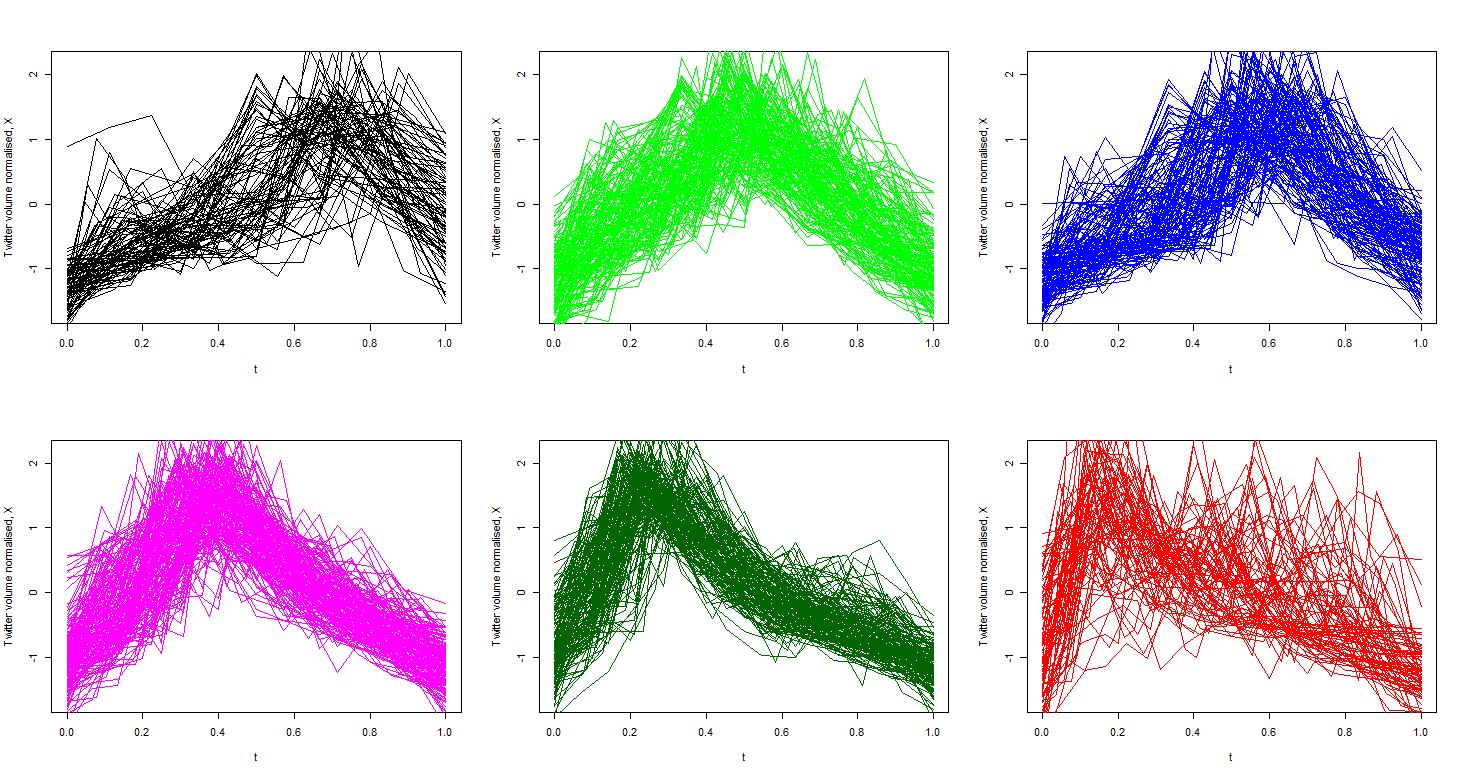}
\label{fig_second_case}}
\caption{
\label{fig_sim}%
Different clusters of Twitter sentiment bursts using (a ) K-means with Euclidean distance, (b) K-means with DTW and (c) K-means Slopes-based approach.}
\end{figure*}

K-Means clustering of Twitter sentiment events (Fig. \ref{fig_sim}) using the three methods described in the Methodology section revealed that all three methods produce clusters of events with similar shapes: a cluster with a peak in the middle, a cluster with a long growth signature and a short relaxation signature, a cluster with a short growth signature and a long relaxation signature. We empirically identified the optimum number of clusters to be six, since six clusters allow to capture the three main dynamics of events shapes as well as variations in the slopes of growth/relaxation of the events. As it was shown by \cite{Sornette:2006}, \cite{Yukie:2013} capturing the slopes of growth and relaxation can help to understand the underlying dynamic of the events (endogenous versus exogenous nature).

Even though all three clustering methods captured the three types of Twitter sentiment shapes, the distance among cluster objects or the measure of spread was different for the three approaches:

\begin{itemize}

\item \textbf{Euclidean distance results.} Twitter events extracted from Twitter sentiment time series have a duration in the range between 7 and 42 days. Since Euclidean distance performs linear mapping it failed to match together events that have different length but similar shape of growth and relaxation (Fig. \ref{fig_sim}(a)). For example, let's consider three different events: event one has a duration of twenty days and sharp growth and relaxation signatures; event two has similar to the first event growth/relaxation signatures, but lasts 40 days; event three lasts 17 days and has a wider shape (the slopes of growth/relaxation are not as sharp as in the first and second events). For the purpose of our analysis we would like to see events one and two clustered together since they have similar shapes (growth/relaxation signatures), however, using the Euclidean distance events one and three will be matched into one cluster since they have similar length and the Euclidean distance will be smaller than the distance between the first and the second events. Apart from that Euclidean distance doesn't handle outliers, and it is very sensitive in respect to signal transformations: shifting, amplitude and time scaling. These drawbacks make Euclidean inappropriate for our application.

\item \textbf{DTW results.} DTW was designed to handle time sequences of varying length and match shapes that do not line up in X-axis by using non-linear mapping. This approach will allow to match together events one and two from the previous example. However, the non-linear mapping introduces a different challenge for clustering: in cases when two events have similar slopes of growth/relaxation, but the location of the peaks differs in time (for example, event one has a short growth and event two has a long growth) the two events will be clustered together (see the event type number four of magenta color in the Fig. \ref{fig_sim}(b)). We believe, that the clustering should not only consider the slopes of the growth/relaxation, but also the location of the peak in time. Another draw-back of DTW is it's computational complexity. 

\item \textbf{Euclidean slopes distance results.} Our new ESD approach, based on the slopes of the strips, produced the most clean results. Fig. \ref{fig_sim}(c)) shows that the distance between time-series within each class is smaller than it is for the other two clustering methods. Our approach allows to solve the problems faced with the other methods:

\begin{enumerate}
\item ESD allows to cluster time-series of different length by automatically normalising them to a number of data points equal to the number of stripes. The clustering considers the growth/relaxation signatures of the events, since the higher level feature of the first derivative allows to extract information about the shape. This solves the problem experienced with the Euclidean distance.
\item ESD approach is using linear mapping between data points which allows to capture the location of the peak in time and solves the problem of DTW.
\item By taking into consideration only high-level features of the time-series (slopes) we reduce dimensionality and thus, reduce noise. This also significantly reduces the calculation time. 
\end{enumerate}

\end{itemize}

These results showed that our clustering approach based on the Euclidean Slopes Distance produced superior results. In the further analysis we only used ESD distance for clustering Twitter events. All further mentioning of the different clusters of event will refer to the clusters produce by ESD method (Fig. \ref{fig_sim}(c)).

\subsection{Statistics of predicted sales events based on Twitter sentiment.}

As described in the Methodology section we implemented two method: in the first method we analysed the aggregated Twitter signal and it's power to predict sales events, in the second method we clustered Twitter events into six classes and analysed the predictive power of each Twitter class independently.

\subsubsection{Method 1: Predictive power analysis of the aggregated Twitter signal.}

According to the method described in subsection 5.4 we calculate a cumulative probability of having a sales event after a Twitter sentiment event.

\begin{figure*}[t!]
  \centering
	\subfloat[]{\includegraphics[width=2.2in]{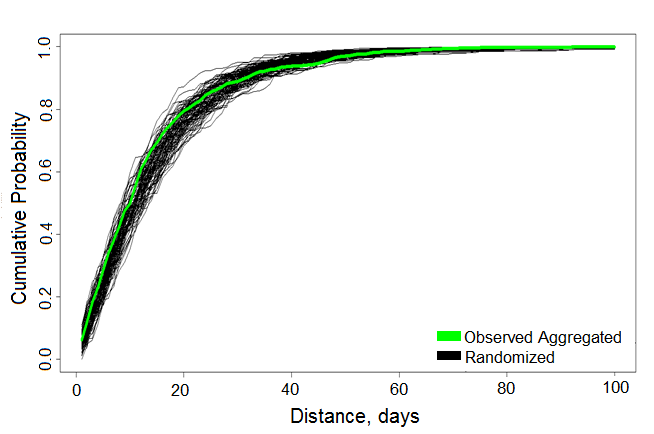}
\label{fig_CP_first_case}}
\hfil
\subfloat[]{\includegraphics[width=2.2in]{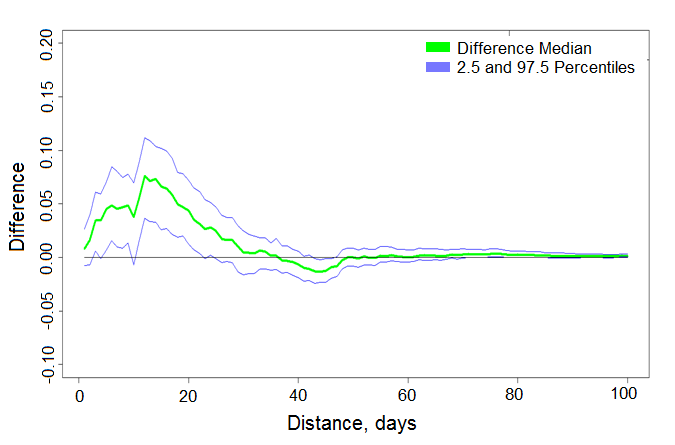}
\label{fig_CP_second_case}}
\caption{(a) Cumulative probability of having a sale event within a specified time interval after a Twitter Sentiment event, (b) Difference between cumulative probabilities of observed and randomised data for Twitter Sentiment. The green line represents empirical results, the black lines represent randomised results and the blue lines represent 95\% confidence interval.}
\label{fig_cumProbab}
\end{figure*}

In figure \ref{fig_cumProbab} (a) the green graph is the empirical cumulative probability and the black graphs are cumulative probabilities for the randomised events sequences. We can observe that the green graph is above the majority of the black graphs for 3 weeks after the Twitter event. In order to quantify the difference between empirical and randomised results we calculate and plot the difference along with 95\% confidence interval.

Figure \ref{fig_cumProbab} (b) shows, that the difference between the empirical and randomised results is significant within a confidence interval of 95\%. According to these results we can reject the null hypothesis and conclude that sales events follow Twitter events in a non-random manner. 

\begin{table}[h]
\caption{Number of successful events for empirical and randomised sequences from when the total number of events is equal to 810.}
\centering
\begin{tabular}{P{1.2in}|P{1.6in}|P{1.6in}}  
 \textbf{Sentiment events} & \textbf{Predictive within 7 days} & \textbf{Predictive within 21 days}  \\ \hline 
\textbf{Empirical Results} & 161 & 434  \\ \hline 
\textbf{randomised Results 95\% CI} & 136.93 [134.9; 138.96] & 417.52 [415.09; 419.96] \\ 
\end{tabular}
\label{tb:sucEvents}
\end{table}

As a second test of significance we look at the statistics of successful events in accordance with the method described in the Methodology section (Fig. \ref{fig:sucSpikes}). Table \ref{tb:sucEvents} shows the number of Twitter sentiment events that have a sale following them within 7 and 21 days. The results are computed for empirical results and for randomised sales sequences. For both 7 and 21 days the number of empirically observed successful events is significantly larger than the number of successful events when sales were randomised. For example, we empirically observed that 434 Twitter sentiment events had at least one sale event following them within 21 days. Conversely, in case when the positions of sales events were randomised, we observed that the average number of Twitter events that had a sales event following them was 417.52 with the 95\% CI [415.09; 419.96]. We therefore observe that the empirical result of 434 is outside of the 95\% CI suggesting that the cases when Twitter events precede sales events do not happen by chance.

\subsubsection{Method 2: Predictive power analysis of different clusters of Twitter events.}

In the second method we used clustering mechanism described in the section 5.3. We obtained six clusters of Twitter Sentiment events (Fig. \ref{fig_sim} (c)).

\begin{figure*}[t!]
  \centering
	\includegraphics[width=4.9in]{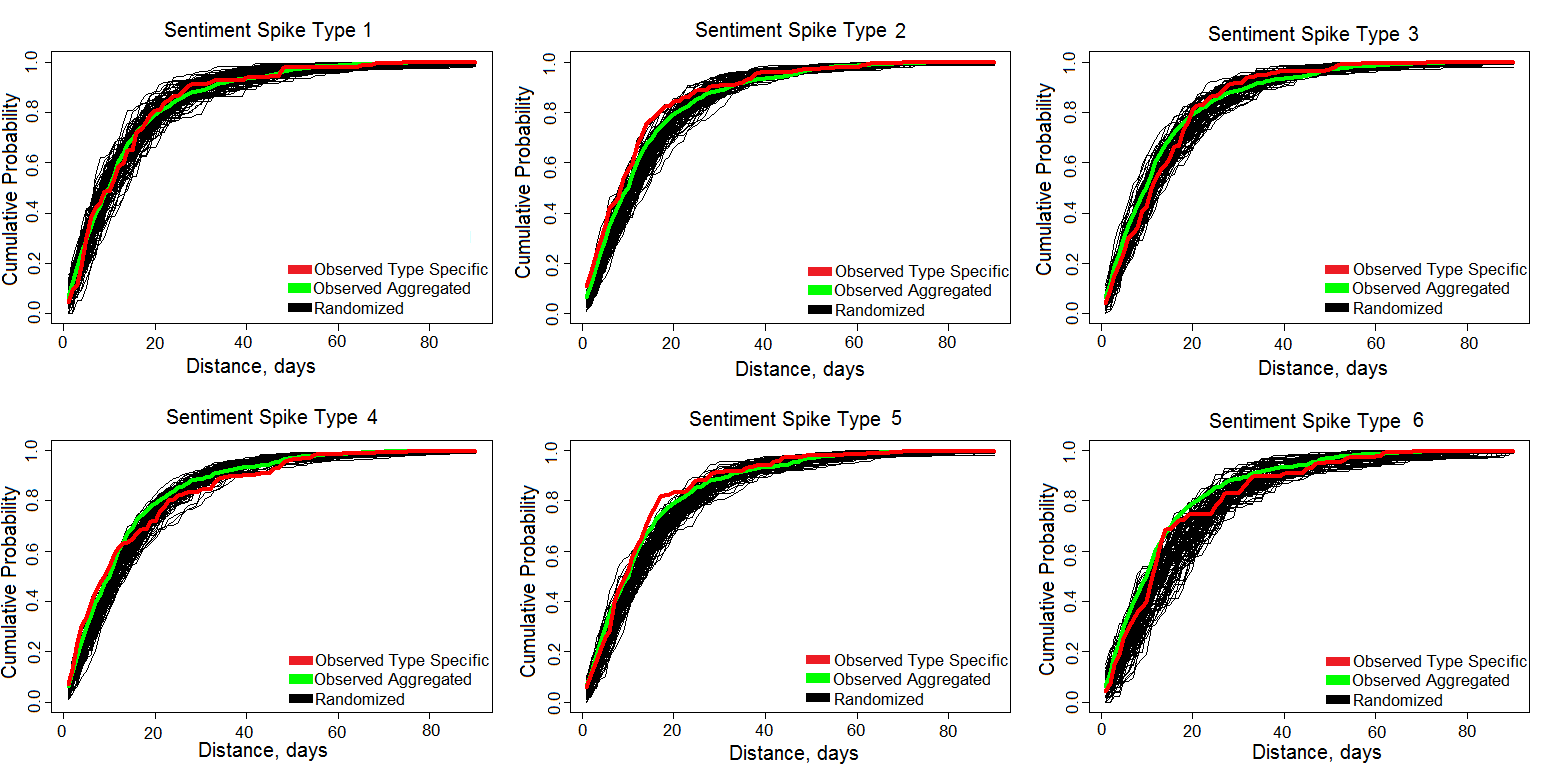}
\caption{\label{fig:SentCumProab_Types}Cumulative probability to have a sale event after Twitter Sentiment events of different types. The red lines correspond to the cumulative probability of specific Twitter type, the green lines correspond to the cumulative probability of the aggregated Twitter signal, the black lines are the cumulative probabilities when sales positions are randomised. Different clusters were obtained using Euclidean Slopes Distance approach.}
\end{figure*}

\begin{figure*}[t!]
  \centering
	\includegraphics[width=4.9in]{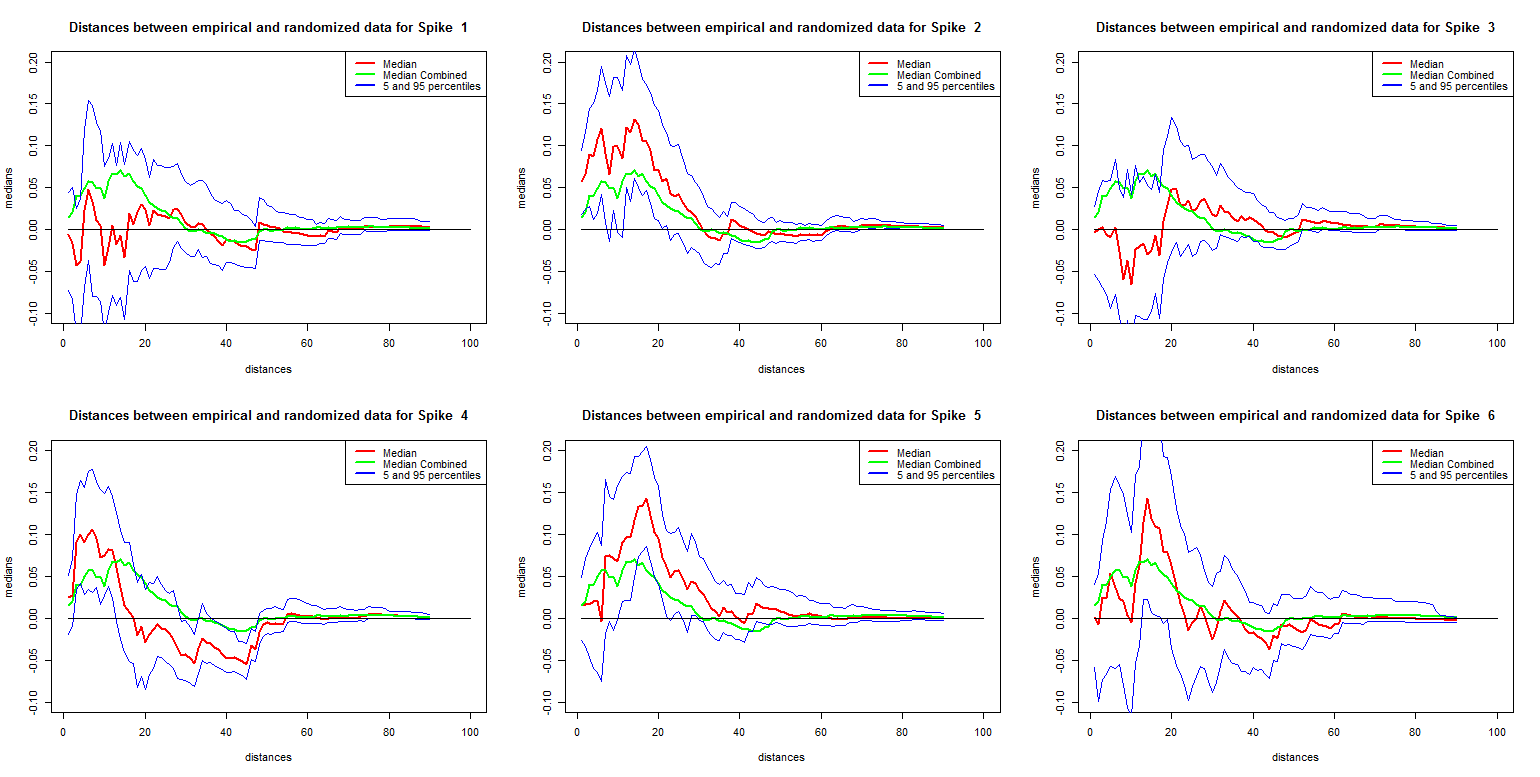}
\caption{\label{fig:SentDif_Types}Difference between cumulative probabilities of observed and randomised data for Twitter Sentiment events of different types. The red lines correspond to the difference for the specific Twitter type, the green lines correspond to the difference for the aggregated Twitter signal, the blue lines represent the 95\% confidence interval. Different clusters were obtained using Euclidean Slopes Distance approach.}
\end{figure*}

Like for the first method we calculate the cumulative probability of having at least one sale event after the Twitter event, however, in this case the probability is computed individually for every Twitter event. We obtain six graphs of cumulative probabilities. To compare the predictive power of each individual Twitter cluster we plot the cumulative probability of the aggregated signal along with the results of individual event types (Fig. \ref{fig:SentCumProab_Types}). We observe that for some events types their cumulative probability is higher than the cumulative probability of the aggregated Twitter Sentiment signal for some periods of time. Plotting the difference between the empirical and randomised results allows us to gain a better understanding of the dynamic (Fig.  \ref{fig:SentDif_Types}).

For events types 2, 4, 5 and 6 we can observe that: 1) the lower border of 95\% confidence interval is above zero and 2) the difference between observed cumulative probability for a specific Twitter type and the randomised sequences (red graph) is greater than the difference between the observed cumulative probability for the aggregated Twitter signal and the randomised data (green graph). The first result indicates sales events follow Twitter sentiment events of type 2, 4, 5 and 6 in a non-random manner. The second result means that the signal of the event types 2, 4, 5 and 6 has a better predictive power than the aggregated Twitter signal. This is a very important finding which means that by using the signals of event types 2, 4, 5 and 6 instead of the aggregated Twitter signal, we can achieve higher accuracy of predicting sales events. It is interesting to notice that event type 2 has a consistent significant predictive power during the first 3 weeks after the event (the 95\% CI is continuously above zero); while event type 4 shows significant predictive power only during the first two weeks, event type 5 is predictive between 12 and 22 days, and event type 6 is only predictive for a short period of time during the third week after the event. 

This information can be incorporated to build a forecasting model that will consider the predictive power of different Twitter events at different distances. For example, the events of Twitter Sentiment of type 2 can be used to predict sales events within the first 3 weeks after the twitter event whereas Twitter events of type 4 can be used to predict sales events only within the first 2 weeks after a Twitter event.

We also tested the hypothesis that the type of Twitter sentiment events that appear before sales events is random. For this purpose we compared the fraction of successful Twitter sentiment events of different types to the fraction of all (successful and unsuccessful) Twitter sentiment events of different types (the notion of the successful Twitter event is explained in the section 5).

\begin{table}[h]
\centering
\begin{tabular}{P{0.9in}|P{0.6in}|P{0.6in}|P{0.6in}|P{0.6in}|P{0.6in}|P{0.5in}}  
 \textbf{Spike Types} & \textbf{Spike type 1} & \textbf{Spike type 2} & \textbf{Spike type 3} & \textbf{Spike type 4} & \textbf{Spike type 5} & \textbf{Spike type 6} \\ \hline 
\textbf{Observed, \%} & 10.55 & 23.6 & 13.66 & 26.09 & 16.77 & 9.32 \\ \hline 
\textbf{Random 95\% CI} & 10.88 [10.43; 11.32] & 20.22 [19.56; 20.86] & 16.91 [16.28; 17.54] & 22.21 [21.6; 23.12] & 20.18 [19.31; 20.71] &  9.62 [9.18; 10.05]\\ 
\end{tabular}
\caption{\label{tb:proportions_7days} Relative frequencies of successful Twitter spikes of different types. Success criteria: sales spike happens within 7 days.}
\end{table}

\begin{table}[h]
\centering
\begin{tabular}{P{0.9in}|P{0.6in}|P{0.6in}|P{0.6in}|P{0.6in}|P{0.6in}|P{0.5in}}  
 \textbf{Spike Types} & \textbf{Spike type 1} & \textbf{Spike type 2} & \textbf{Spike type 3} & \textbf{Spike type 4} & \textbf{Spike type 5} & \textbf{Spike type 6} \\ \hline 
\textbf{Observed, \%} & 11.06 & 20.51 & 17.74 & 22.35 & 19.35 & 8.99 \\ \hline 
\textbf{Random 95\% CI} & 10.91 [10.61; 11.08] & 20.30 [20.07; 20.53] & 16.43 [16.23; 17.14] & 22.63 [22.38; 23.13] & 19.90 [19.38; 20.13] &  9.83 [9.61; 10.00]\\ 
\end{tabular}
\caption{\label{tb:proportions_21days} Relative frequencies of successful Twitter spikes of different types. Success criteria: sales spike happens within 21 days.}
\end{table}

Results are shown in table \ref{tb:proportions_7days}, where  the first line of the table presents the proportions of different events classes for 161 Twitter events that were successful within 7 days. To test our null hypothesis we apply randomisation test: randomising positions of sales events, identifying successful Twitter events and calculating proportional representation of different types for the randomised sales. We repeat randomisation test 1000 times and then calculate the average values with 95\% confidence intervals (the third row of the table \ref{tb:proportions_7days}).

Comparing the observed relative frequencies (the first row) with the proportions after randomisation (the second row) we see that the results for events types 2, 3, 4 and 5 are significantly different from random results. The events types 3 and 5 are significantly unrepresented while events types 2 and 4 are significantly overrepresented. 

Very similar results can be observed for the 434 events that were successful within 21 day (table \ref{tb:proportions_21days}). We see that the relative frequencies of types 3, 4, 5 and 6 are significantly different from random.

We made the analysis for the distance of 7 and 21 days after the Twitter event, because according to the results in Fig. \ref{fig:SentDif_Types} some events have predictive power up to 7 days after the event, however all of them loose their predictive power around the end of the third week after the Twitter event - 21 days.  Comparing the results for different distances - 7 and 21 days, we observe that relative frequencies of events types changes depending on the time interval. For example, the event of type 4 is overrepresented for the 7 days distance and unrepresented for the 21 days distance. This result is consistent with the analysis of cumulative probabilities (Fig. \ref{fig:SentDif_Types}). Indeed, we see that events of type 4 have significant predictive power during the first 1-2 weeks after the event and no predictive power on the day 21 and later. 

Since we observed significant deviations from random in the success frequencies of different types of events we can reject the null hypothesis and conclude, that Twitter events contain information structure and the occurrence of different Twitter events types before sales events is not random. 

We performed the same statistical validation for the classes of events obtained through the other two clustering methods: based on the Eucledian distance and based on DTW. The results are presented in the Appendix A. The clusters obtained through DTW clustering do not reveal any significant predictive power. The results based on the Euclidean distance reveal some clusters of events that have ability to predict sales events better than random, however the number of days for which the results are significant is less than the results of the clusters obtained through the slopes based clustering. This observation confirms that the method of slopes-based clustering is the most suitable for this type of problem comparing to other existing methods.

\section{Conclusion}

We studied the relationship between Twitter events and events in consumer sales, analysing sales for 75 brands and 150 million tweets, along with their sentiment.  For the first time such analysis is conducted on a large scale for the consumer products that do no receive large social attention. 

Our research presents a contribution to the field of the event study. We suggest a new approach for events detection that considers the duration of the Twitter event. We argue, that information in Twitter spreads over time and taking in account only the peak day is not sufficient for the accurate definition of the event. We also suggest a novel approach of clustering events into different types based on the  dynamic of information propagation. We highlight, that such features of the event as the location of the event's peak, it's growth and relaxation signatures, contain valuable information and must be used in the analytics.

The  predictive power of Twitter events in this study was evaluated using two scenarios: in the first scenario we perform the analysis for the aggregated Twitter signal considering only tweets' sentiment; in the second scenario we cluster Twitter sentiment events based on their shape and then calculate the statistics of successful predictions separately for every events type. The results of our analysis can be summarised as follows:

\begin{itemize}
\item Aggregated Twitter events can be used to predict sales spikes.
\item Spikes can be separated into categories based on their shapes (position of the peak, growth and relaxation signatures).
\item Different spikes shapes are differently associated with sales.
\item Some spikes types have a higher predictive power than the aggregated Twitter signal.
\end{itemize}

To summarise, this paper fills the gap in social-media events study for predicting events in sales. We contribute to the field of the event study by proposing a different process of identifying events, which is more suitable for the modern social-media landscape. We also contribute to the field of forecasting by providing an extensive study of a large dataset, covering 75 brands and over 150 million Twitter messages. We show that events in social-media can be classified based on their growth and relaxation signatures. The important result of our research is identification of events types that have a power to predict events in sales and this predictive power is more significant than the aggregated Twitter signal. As the future direction of our research we aim to understand what different events' shapes represent in terms of Twitter dynamics and content (persistence of news, importance) and plan to incorporate the extracted knowledge into a forecasting model for consumer sales.

\section*{Acknowledgments}

This work was supported by the company Certona Corporation. T.T.P.S. acknowledges financial support from CNPq - The Brazilian National Council for Scientific and Technological Development.

\newpage
\clearpage
\section{Appendix A}

\begin{figure} 
\centering
\includegraphics[width=0.8\textwidth]{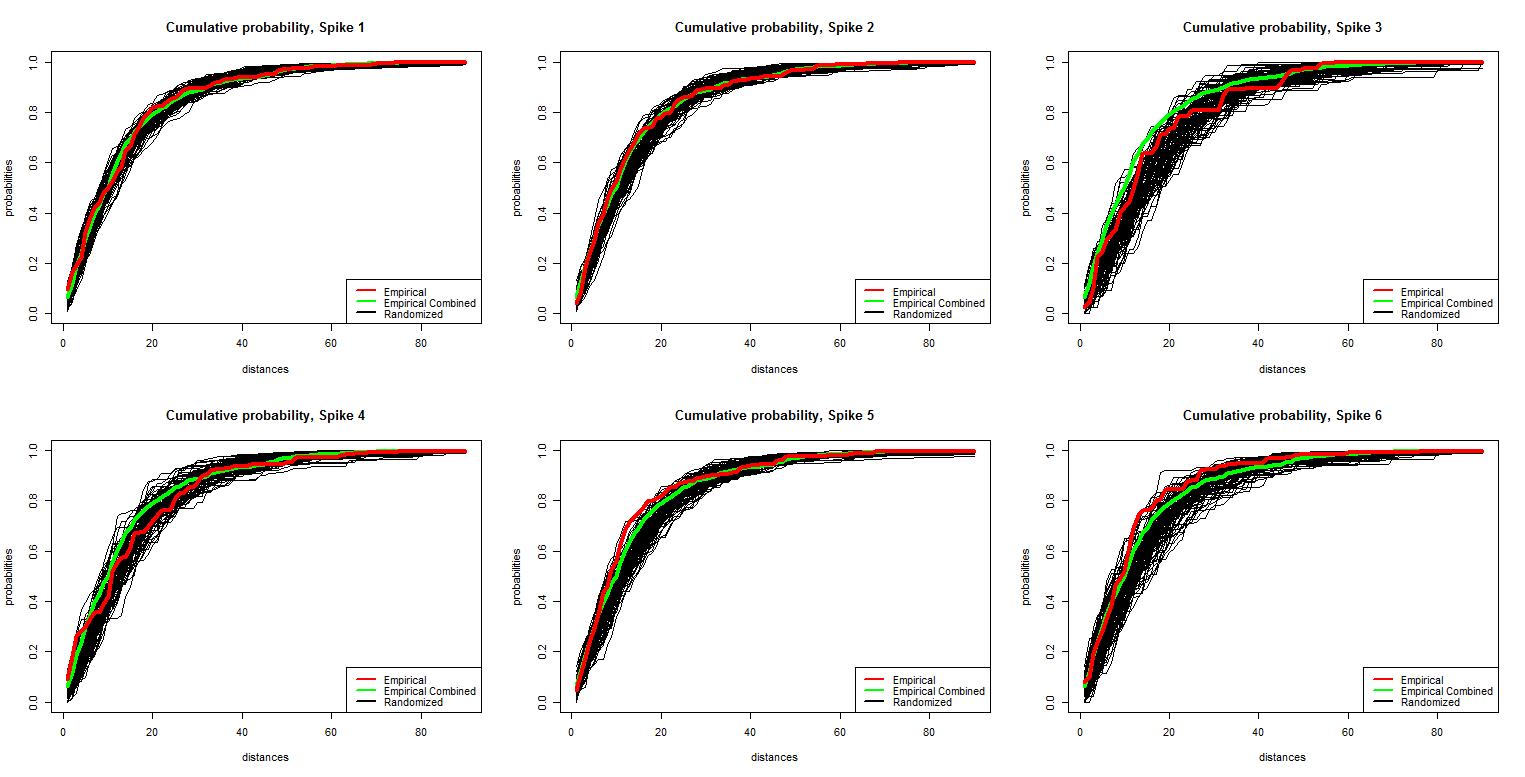}
\caption{\label{fig:cumProbab_EuclDist} Cumulative probability to have a sale event after a Twitter event using Euclidean distance based clustering (red lines - specific Twitter type; green lines - aggregated Twitter signal; black lines - the case of randomised sales.}
\end{figure}

\begin{figure}
\centering
\includegraphics[width=0.8\textwidth]{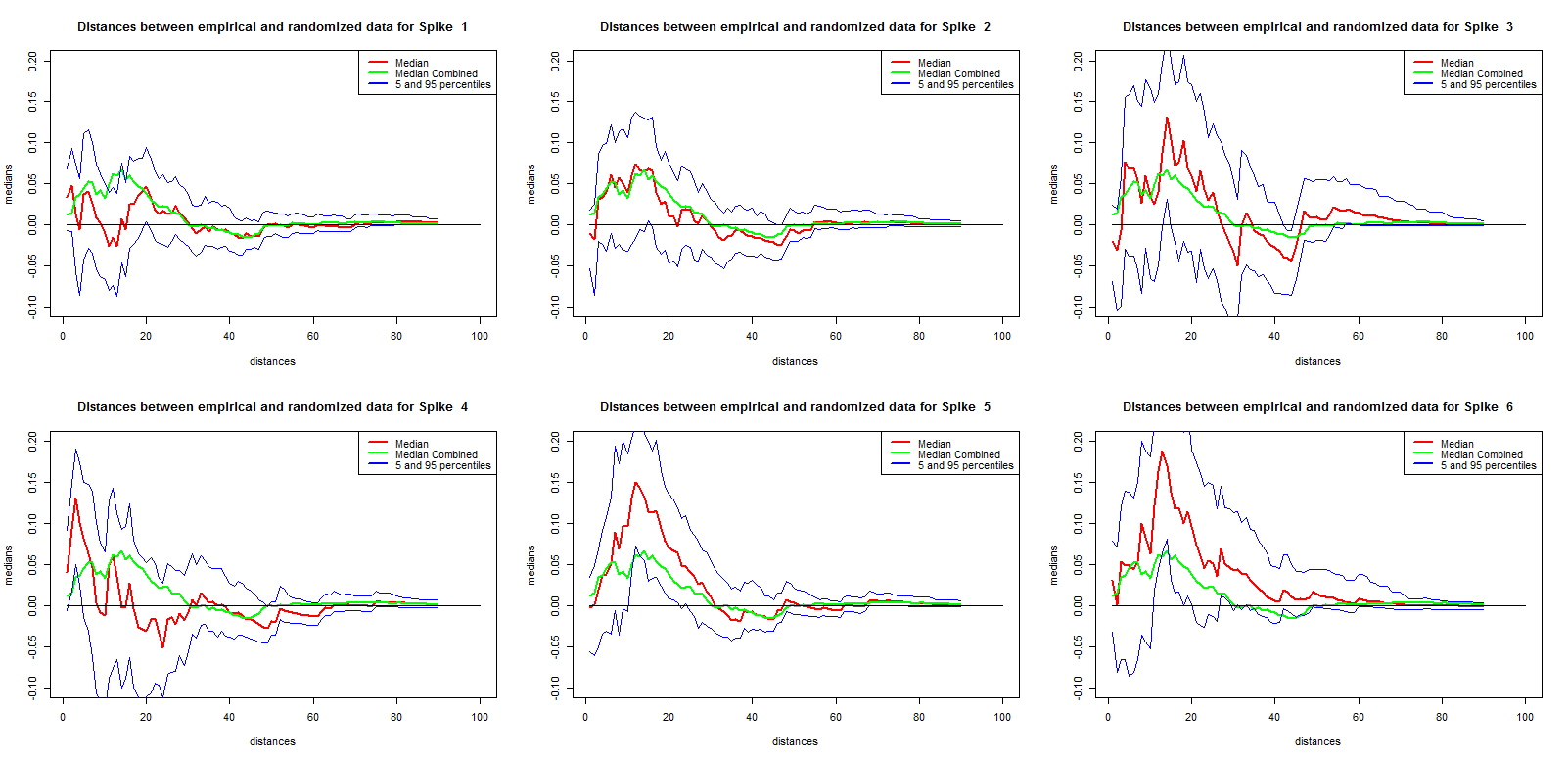}
\caption{\label{fig:diff_EuclDist} Difference between cumulative probabilities of observed and randomised data using Euclidean distance based clustering (red lines - specific Twitter type; green lines - aggregated Twitter signal; blue lines - the 95\% confidence interval).}
\end{figure}

\begin{figure}
\centering
\includegraphics[width=0.8\textwidth]{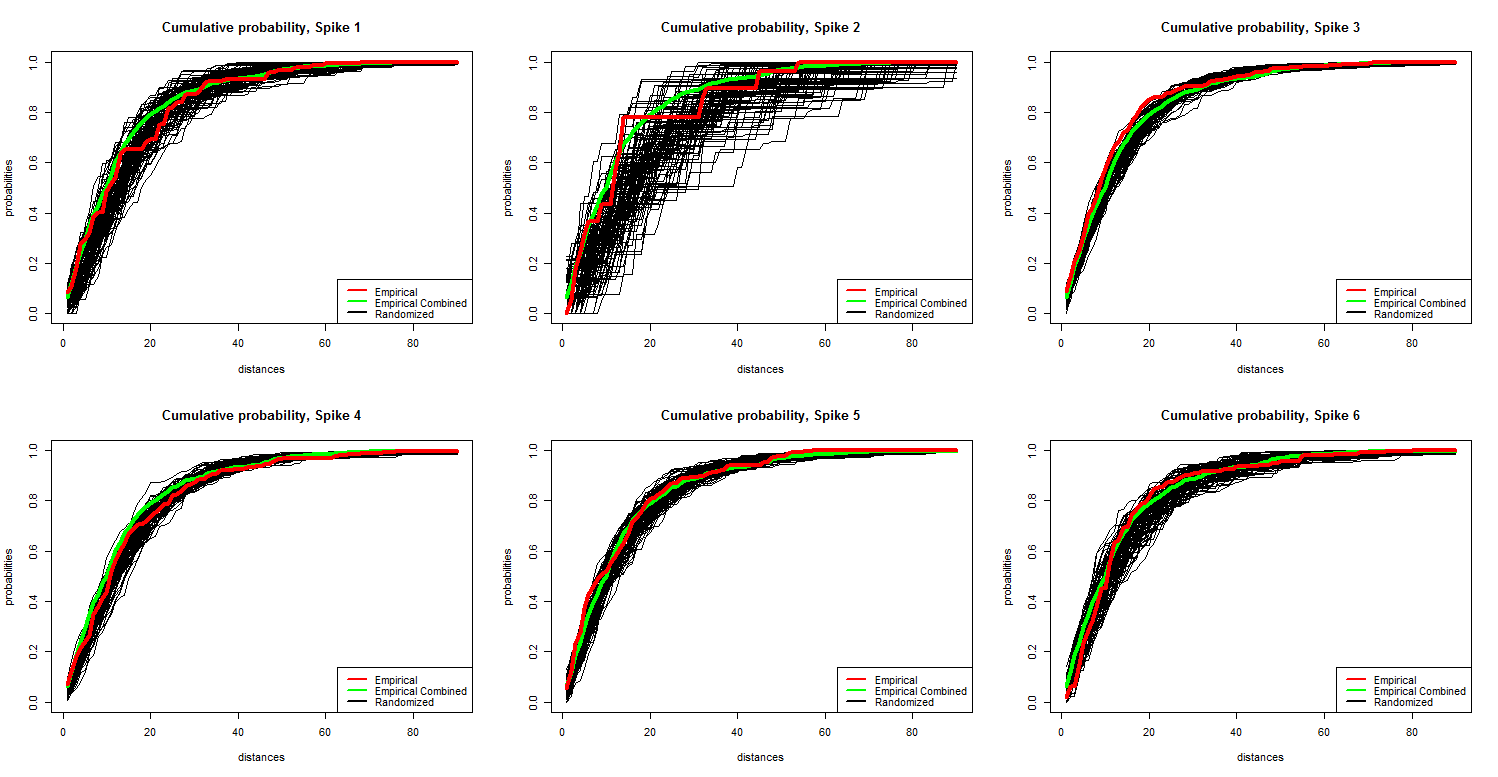}
\caption{\label{fig:cumProbab_DTW} Cumulative probability to have a sale event after a Twitter event using Dynamic Time Warping based clustering (red lines - specific Twitter type; green lines - aggregated Twitter signal; black lines - the case of randomised sales).}
\end{figure}

\begin{figure}
\centering
\includegraphics[width=0.8\textwidth]{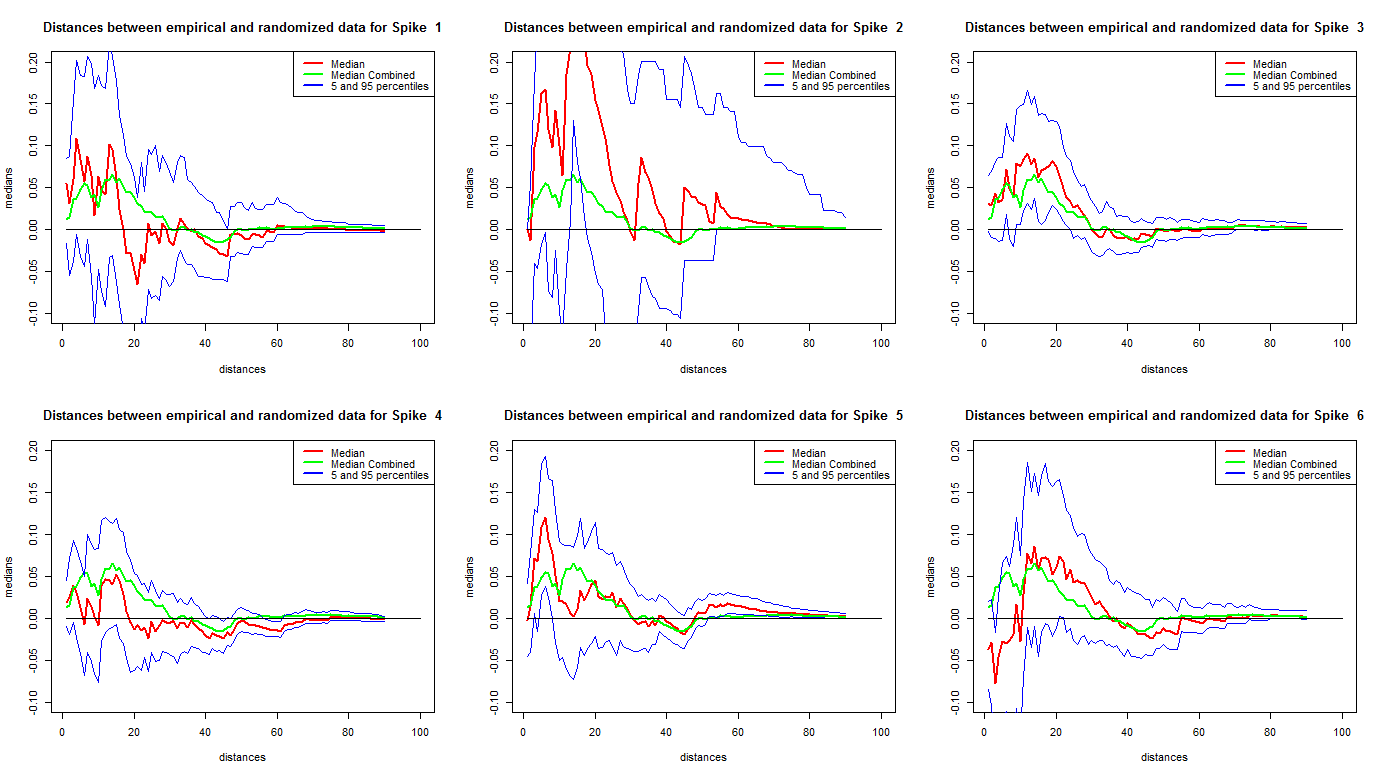}
\caption{\label{fig:diff_DTW} Difference between cumulative probabilities of observed and randomised data using Dynamic Time Warping based clustering (red lines - specific Twitter type; green lines - aggregated Twitter signal; blue lines - the 95\% confidence interval).}
\end{figure}

\newpage
\clearpage

\bibliographystyle{IEEEtran}
\bibliography{chapter}

\end{document}